\documentstyle[12pt]{report}
\begin{document}
\baselineskip=20pt

\begin{center}
{\bf MULTISCALING TO STANDARD SCALING CROSSOVER}\\
{\bf IN THE BRAY-HUMAYUN MODEL}\\
{\bf FOR PHASE ORDERING KINETICS}\\
{}~~\\
{}~~\\
{}~~\\
C. Castellano$^{1}$ and M. Zannetti$^{2}$\\
{}~~\\
{\small \it $1$ Dipartimento di Scienze Fisiche, Universit\`{a} di Napoli,
Mostra d'Oltremare Pad.19, 80125 Napoli, Italy\\
$2$ Istituto Nazionale di Fisica della Materia, Unit\`{a} di Salerno\\
and  Dipartimento di Fisica, Universit\`{a} di Salerno,
84081 Baronissi (SA), Italy}
\end{center}
{}~~\\
\begin{center}
{\bf Abstract}
\end{center}
\begin{quote}
The Bray-Humayun model for phase ordering dynamics
is solved numerically in one and two
space dimensions with conserved and non conserved order
parameter. The scaling properties are analysed in detail
finding the crossover from multiscaling to standard scaling
in the conserved case. Both in the nonconserved case
and in the conserved case when standard scaling holds
the novel feature of an exponential tail in the scaling
function is found.
\end{quote}

{}~~\\
05.70.Fh 64.60.Cn 64.60.My 64.75.+g

\newpage

\setcounter{chapter}{1}
\setcounter{equation}{0}

\section*{1 - Introduction}

\vspace{5mm}

In this paper we are concerned with scaling behaviour in
the phase ordering dynamics of a system quenched below the
critical point[1]. Specifically, we consider a
system with an $N$-component order parameter
$\vec{\phi}(\vec{x})=(\phi_{1}(\vec{x}),...,\phi_{N}(\vec{x}))$
quenched from high temperature to zero temperature whose
dynamics are described by the zero noise Langevin equation
\begin{equation}
\frac{\partial \vec{\phi}(\vec{x},t)}{\partial t} =
(i \nabla)^{p} \left [ \nabla ^{2} \vec{\phi} -
\frac{\partial V(\vec{\phi})}{\partial \vec{\phi}} \right ]
\end{equation}
where $p=0$ for non conserved order parameter (NCOP), $p=2$
for conserved order parameter (COP) and
$V(\vec{\phi}) = \frac{r}{2} \vec{\phi}^{2} +
\frac{g}{4N} (\vec{\phi}^{2})^{2}$
is the local potential with $(r<0,g>0)$.
One of the reasons for the continuing interest in this
type of problem is that a theoretical derivation of scaling
on a first principles basis is still lacking except for
a few exactly soluble models[2,3].

Let us first give a qualitative description of what goes on
during the phase ordering process. Initially the system is prepared
in a high temperature state where the order parameter is spatially
uncorrelated
\begin{equation}
<\phi_{\alpha}(\vec{x},0)\phi_{\beta}(\vec{x}^{\prime},0)>
=\Delta \delta_{\alpha \beta}(\vec{x}-\vec{x}^{\prime})
\end{equation}
with $\alpha,\beta=1,...,N$ and $\Delta$ is a constant. The local
order parameter probability distribution has
as a peak of width $\Delta$ centered about $\phi_{\alpha}=0$.
As the quench develops there is first a fast process (early
stage) where this probability distribution, after
a short time $t_{0}$, relaxes to equilibrium in the local potential,
depleting the origin and developing a peak structure all
around the bottom of the potential.
In this span of time the local variance
$<\phi_{\alpha}^{2}(\vec{x},t_{0})>=S(t_{0})$ loses
memory of the initial condition $\Delta$ reaching a
value very close to the final saturation value
$S_{eq}= -r/g$. At this point the system
is almost at equilibrium, namely ordered, on a short length scale.
The subsequent time evolution amounts to coarsening of the
ordered regions in order to reduce the excess interfacial free
energy. During this process (late stage) the only
important time dependence is in the linear size of the
ordered regions which typically grows according to a
power law $L(t) \sim t^{1/z}$ with $z=2$ for NCOP and
$z=3$ or $z=4$ for COP respectively with $N=1$ or $N>1$.
When $L(t)$ is large enough to dominate all other lengths
the quantities of interest exhibit scaling. The
main observables are the equal time order parameter
correlation function
$G(\vec{r},t)= <\phi_{\alpha}(\vec{x},t)
\phi_{\alpha}(\vec{x}+\vec{r},t)>$ and the structure factor
$C(\vec{k},t)$ obtained by Fourier transforming $G(\vec{r},t)$
with respect to space. The local variance of the order
parameter is related to these quantities by
$S(t)=G(\vec{r}=0,t)=\int \frac{d \vec{k}}{(2\pi)^{d}}
C(\vec{k},t)$. Actually, we will be interested in the
quantity
\begin{equation}
R(t)=r+gS(t)= g[S(t)-S_{eq}]
\end{equation}
which monitors how the saturation value of $S(t)$ is reached.

According to the scaling hypothesis, all time dependence
can be expressed through $L(t)$. The dominant behaviours for
large
$L(t)$ are given by
\begin{equation}
R(t)= -\frac{b}{L^{\theta}(t)}
\end{equation}
with $\theta =2$ for systems with vector order parameter[4] and
\begin{equation}
C(\vec{k},t)=S_{eq} f(L(t),kL(t))
\end{equation}
where the function $f(L,kL)$ must go over to $\delta (k)$
in the limit $L \rightarrow \infty$ in order to reproduce the
Bragg peak corresponding to the final ordered state. In other words
$f(L,kL)$ is a smoothed out $\delta$-function on the scale
$L(t)$ with the normalization property
\begin{equation}
\int \frac{d \vec{k}}{(2\pi)^{d}} f(L(t),kL(t)) =1.
\end{equation}
Experiments, numerical simulations and soluble models,
with the exception of the exact solution of the large-$N$
model with COP[3], yield the following form of scaling
\begin{equation}
f(L,kL) = L^{d} F(kL)
\end{equation}
which we refer to as standard scaling. By contrast, when the model
with $N=\infty$ and COP is solved one finds out that this
pattern of scaling is not obeyed. In that case the system
behaves differently since next to $L(t)$ there is another divergent
length $k_{m}^{-1}(t) \sim L/(\ln L)^{1/4}$ where $k_{m}(t)$
is the peak wave vector of the structure factor. Consequently,
there appears a logarithmic correction in (1.4)
\begin{equation}
R(t)=-\frac{b}{L^{2}(t)}(\ln L)^{1/2}
\end{equation}
and in place of (1.7) one has the qualitatively different form
\begin{equation}
f(L,k/k_{m}) \sim (L^{2}k_{m}^{2-d})^{\psi(k/k_{m})}
\end{equation}
with $\psi(x)=1-(x^{2}-1)^{2}$. This pattern of scaling is
referred to as multiscaling.

It should be stressed that the
above solution of the $N=\infty$ model is the only available
analytical solution of a system with COP. Hence, due to the
difficulty of discriminating between multiscaling and standard
scaling on the basis of the usual data collapse analysis, one may
reasonably ask the question whether multiscaling might in fact be
a generic feature of all systems with COP. In other words,
putting $x=kL$ and neglecting logarithmic differences between
$L$ and $k_{m}^{-1}$, i.e. letting $L = u/k_{m}$ with $u$ constant,
one can postulate the general scaling form
\begin{equation}
f(L,x) = \left [ L(t) \right ] ^{\varphi(x)} F(x)
\end{equation}
which contains (1.7) and (1.9) as particular cases respectively
with $\varphi(x) \equiv d$ and $\varphi(x) =d \psi(x/u)$.
It is then matter of computation or experiment
to extract the spectrum of exponents $\varphi(x)$ and to check
whether it is flat like in standard scaling or there is a
genuine $x$-dependence implying multiscaling.

This kind of analysis has been carried out on the data for
the structure factor obtained from the simulation[5,6,7] of systems
with COP and with $N$ ranging from $1$ to $4$ in two and
three dimensions. In all of these cases the observed behaviour of
$\varphi(x)$ is consistent with the flat spectrum characteristic
of standard scaling. Furthermore, analytical work of Bray
and Humayun[8] (BH) on a model with $N$ large but finite and
$d>1$ suggests that standard scaling holds for any finite $N$,
while multiscaling is only a feature of the special case
$N=\infty$. Actually, the picture that BH put forward is that
there exists a crossover time $t^{\ast}$
which depends on $N$ and the initial condition $\Delta$
and such that multiscaling holds for $t<t^{\ast}$
whilst for $t>t^{\ast}$ standard scaling sets in.
Therefore, different asymptotic behaviours are obtained
according to the order of the limits $t \rightarrow \infty$
and $N \rightarrow \infty$, as it was conjectured very
early on by Yoshi Oono[9].
If the limit $t \rightarrow \infty$
is taken first, standard scaling is observed asymptotically
for any value of $N$, as long as $N < \infty$. Conversely,
if the limit $N \rightarrow \infty$ is taken first, then the
crossover time $t^{\ast}$ diverges and the asymptotic behaviour
exhibits multiscaling since the regime of standard scaling
can never be reached.

What is at stake in this question of standard scaling vs.
multiscaling is the nature of the symmetry underlying scaling
behaviour[5]. The results of the simulations could be
regarded as non conclusive, since one could well imagine a spectrum
$\varphi(x)$ which is dependent on $N$ and which interpolates
between the $N=\infty$ behaviour and the standard scaling
behaviour as $N$ gets smaller and smaller. Then, for values
of $N$ of order unity, such as in the simulations, it might
be difficult to decide whether a $\varphi(x)$ with a weak
dependence on $x$ is evidence for standard scaling or for multiscaling.
Instead, the result of BH is clearcut and states that the symmetry
underlying asymptotic dynamics leads to standard scaling for
any finite $N$.
This result is quite important from the point of view of theory since
theoretical progress in this field so far has heavily relied
on the use of very clever but uncontrolled approximations[10]. This
is due to the difficulty of developing systematic and controlled
approximation schemes. An exception is the $1/N$-expansion
for systems with NCOP[11]. The result of BH eliminates
the possibility of extending the $1/N$-expansion as a systematic expansion
scheme
to the conserved case.

For the relevance of this issue, in this paper we have made a
detailed study of the crossover from multiscaling to standard
scaling through a comparative analysis of the numerical solution
of the BH model with NCOP and with COP.
Our aim is to proceed to an unbiased analysis
of the scaling properties in order to have a check on the BH picture
without any a priori assumption on the type of scaling, and
to analyse in detail the difference between the conserved and
non conserved case.
In this respect our work is quite different from that of Rojas and Bray[12].
These authors do perfom a numerical solution
of the BH model {\it after} the standard scaling ansatz has been made,
while we first solve for the structure factor and then we proceed
to the scaling analysis on the basis of the uncommitted general
form (1.10).

The paper is organized as follows:  in section 2 we present
the model and we elaborate on the difference between standard
scaling and multiscaling, introducing the observables best
suited to distinguish one from the other. In section 3 we
illustrate the method of solution with a test of its
validity made by comparing numerical data with the analytical
solution in the $N=\infty$ case. In section 4 we present the
results for finite $N$ in one and two dimensions and in section 5
we make some concluding remarks.

\vspace{8mm}

\setcounter{chapter}{2}
\setcounter{equation}{0}

\section*{2 - BH Model, Standard Scaling and Multiscaling}

\vspace{5mm}

By using the gaussian auxiliary field method of Mazenko[13],
BH have derived[8] from (1.1) a closed equation of motion
for the equal time correlation function within the framework
of the $1/N$-expansion. Retaining non linear terms up to first
order in $1/N$ one has
\begin{equation}
\frac{\partial G(\vec{r},t)}{\partial t} = 2(i \nabla)^{p}
\left [ \nabla ^{2} G -R(t)\left (G+\frac{1}{N}G^{3} \right ) \right ]
\end{equation}
where $R(t)$ is a function of time which must be determined
by the equilibrium requirement
\begin{equation}
\lim_{t \rightarrow \infty} G(\vec{r}=0,t)= S_{eq} =-r/g.
\end{equation}
The corresponding equation of motion for the structure factor
is obtained after Fourier transforming with respect to space
variables
\begin{equation}
\frac{ \partial C(\vec{k},t)}{\partial t}= -2k^{p} \left [ k^{2} +
R(t) \right ] C(\vec{k},t) -2\frac{k^{p}}{N} R(t) D(\vec{k},t)
\end{equation}
where $D(\vec{k},t)$ is the Fourier transform of $G^{3}(\vec{r},t)$.
Notice that in the limit $N \rightarrow \infty$  (2.3) reduces
to the equation which has been studied in [3]
\begin{equation}
\frac{ \partial C(\vec{k},t)}{\partial t}= -2k^{p} \left [ k^{2} +
R(t) \right ] C(\vec{k},t).
\end{equation}
In this latter case $R(t)$ is defined self-consistently by (1.3).
We shall retain this definition of $R(t)$ also in the finite
$N$ case since from (2.3) follows that in order to reach
equilibrium $R(t)$ must vanish and with the definition (1.3)
the condition $\lim_{t \rightarrow \infty} R(t) =0$ is an
implementation of the requirement (2.2).

Although (2.1) or (2.3) have been
derived by a truncation procedure based on the $1/N$-expansion,
the solution of the equation is not of first order in $1/N$,
since it contains all orders in $1/N$. Actually, it is not
possible to assess precisely what is the relationship of
this solution with the systematic $1/N$-expansion performed on the basic
equation of motion (1.1). Presumably, it is some kind of infinite
partial resummation intertwined with the uncontrolled approximation
inherent in the use of the gaussian auxiliary field method of
Mazenko[10,13]. Hence, (2.1) or (2.3) should be regarded as the definition
of a model, the BH model, for phase ordering dynamics with an
$N$-component vectorial order parameter which in the $N \rightarrow
\infty$ limit reproduces the usual large-$N$ limit of (1.1) for
the dynamics of the structure factor.

In order to make the scaling analysis of the BH model, let us
integrate formally (2.3) from some instant of time $t_{0}$ onward
\begin{eqnarray}
C(\vec{k},t) & = & C(\vec{k},t_{0}) e^{-2 \left [ k^{2+p}(t-t_{0})+
k^{p}\left ( Q(t)-Q(t_{0}) \right ) \right ]} \nonumber \\
             &   & -2\frac{k^{p}}{N} \int_{t_{0}}^{t}
dt^{\prime} R(t^{\prime})
e^{-2\left [ k^{2+p}(t-t^{\prime})+k^{p}\left ( Q(t)-Q(t^{\prime}) \right )
\right ]} D(\vec{k},t^{\prime})
\end{eqnarray}
where $Q(t)=\int_{0}^{t} dt^{\prime} R(t^{\prime})$.
Choosing $t_{0}$ in the scaling region, according to the standard
scaling hypothesis we have the asymptotic behaviours
\begin{eqnarray}
C(\vec{k},t)& = & S_{eq} L^{d}(t)F(x) \\
R(t) &= & -bL^{-2}(t) \\
Q(t)-Q(t_{0})& = & \left \{ \begin{array}{ll}
                        -2b \log (L(t)/L_{0}) & \mbox{, for NCOP} \\
                        -2b(L^{2}(t)-L_{0}^{2}) & \mbox{, for COP}
                        \end{array}
               \right.
\end{eqnarray}
with $x=kL(t)$, $L(t)=t^{1/(2+p)}$,
$L_{0}=L(t_{0})$ and $b$ is a positive constant to be determined.

Let us first consider the case of NCOP. Performing the above ansatz
on (2.5) we find
\begin{eqnarray}
F(x) =  F(x_{0}) (x/x_{0})^{4b-d} e^{-2(x^{2}-x_{0}^{2})} \nonumber \\
              +S_{eq}^2\frac{4b}{N}
\int_{x_{0}}^{x} \frac{d x^{\prime}}{x^{\prime}}(x/x^{\prime})^{4b-d}
 e^{-2(x^{2}-x^{\prime 2})} {\cal D}(x^{\prime})
\end{eqnarray}
where $x_{0}=kL_{0}$ and
${\cal D}(\vec{x}^{\prime})= \int \frac{d\vec{x}_{1}}{(2 \pi)^{d}}
\frac{d\vec{x}_{2}}{(2 \pi)^{d}}
F(\mid \vec{x}^{\prime}-\vec{x}_{1} \mid)
F(\mid \vec{x}_{1}-\vec{x}_{2} \mid) F(x_{2})$.
The condition (2.2) requires
\begin{equation}
\int \frac{d \vec{x}}{(2 \pi)^{d}} F(x) = 1
\end{equation}
which gives an equation for $b$. Letting $x_{0} \rightarrow 0$
and requiring (2.10) to be satisfied, in the $N=\infty$ case
we obtain $4b-d=0$ and
\begin{equation}
F(x)=F(0)e^{-2x^{2}}
\end{equation}
while for finite $N$ we find $4b-d < 0$.

Let us now go to the COP case. Making the scaling ansatz
into (2.5) with $p=2$ we find
\begin{eqnarray}
F(x) = F(x_{0}) (x_{0}/x)^{d}
e^{-2[(x^{4}-x_{0}^{4})-2b(x^{2}-x_{0}^{2})]} \nonumber \\
+ \frac{8b S_{eq}^{2}}{ Nx^{d}} \int_{x_{0}}^{x} d x^{\prime}
x^{\prime d+1} e^{-2 \left [ (x^{4}-x^{\prime 4}) -2b (x^{2}-
x^{\prime 2}) \right ]} {\cal D}(x^{\prime}).
\end{eqnarray}
Now, if we let again $x_{0} \rightarrow 0$, in the $N = \infty$ case
$F(x)$ vanishes identically and it is not possible anymore
to satisfy (2.10). This is
the breakdown of standard scaling in the large-$N$ limit
which leads to multiscaling[3]. Conversely, if $N$ is kept
finite, (2.12) yields
\begin{equation}
F(x) = \frac{8b S_{eq}^{2}}{ Nx^{d}} \int_{0}^{x} d x^{\prime}
x^{\prime d+1} e^{-2 \left [ (x^{4}-x^{\prime 4}) -2b (x^{2}-
x^{\prime 2}) \right ]} {\cal D}(x^{\prime}).
\end{equation}
This is the equation that BH have solved finding a non trivial
solution and reaching the conclusion
that for any finite $N$ standard scaling holds also for
systems with COP.

According to the above discussion the first term in the right
hand side of (2.5), the one which survives after $N \rightarrow \infty$,
is responsible for multiscaling behaviour while the second
one is responsible for standard scaling. The competition of these two
terms is expected to generate
a crossover time $t^{\ast}$ such that multiscaling behaviour of the
type found with $N=\infty$ holds for $t<t^{\ast}$ while
standard scaling  eventually sets in for $t > t^{\ast}$.

In the following we
will make a numerical study of the scaling properties of the
structure factor in the BH model on the basis of the general scaling
form (1.10). The primary interest is in the
discrimination between standard scaling and multiscaling
and in the study of the crossover. The analysis will be carried out
through the behaviour of the spectrum of exponents $\varphi(x)$
as described in the Introduction and through the behaviour of the quantity
\begin{equation}
Y(t)= -R(t) L^{2}(t)
\end{equation}
which discriminates between standard scaling and multiscaling
on the basis of the asymptotic behaviours
\begin{equation}
Y(t) \left \{ \begin{array} {ll}
                     = b(N) & \mbox{, for standard scaling} \\
                    \sim (\ln t)^{1/2} & \mbox{, for multiscaling.}
                   \end{array}
          \right.
\end{equation}

\vspace{8mm}

\setcounter{chapter}{3}
\setcounter{equation}{0}

\section*{3 - Method of solution and numerical results for $N = \infty$}

\vspace{5mm}

In order to investigate the scaling properties of the BH model
the discretized version of (2.1) is integrated numerically
via a simple finite difference first order Euler scheme.
The initial condition is given by $G(\vec{r},0) = \Delta$ for $\vec{r}=0$ and
$G(\vec{r},0)=0$ elsewhere.
The boundary conditions are chosen to be periodic, but open
conditions have been tested not to affect the final results.
 From the values of $G(\vec{r},t)$ the structure factor is then obtained via
Fast Fourier Transform and from these two functions all quantities of interest
are computed.

Two opposite requirements enter in the choice of the parameters
of the numerical solution, and in particular of the linear dimension $L$ of the
system.
A large number of lattice sites is desirable to avoid
that the discretization of space may
hide the subtle difference between standard scaling and multiscaling.
On the other hand, fewer sites speed up the computation
and investigation of later times is feasible.
Resorting to parallel computing we have managed to perform the numerical
integration on large systems and for sufficiently long times.
In principle, one can solve (2.3) to obtain
directly the structure factor, but the presence in (2.3) of a double
convolution integral, which cannot be parallelized efficiently, makes this
alternative computational scheme much slower.

For all runs the value of the mesh size has been taken
$\Delta x = 1$, while the
time step $\Delta t$ has been changed
depending on the values of $N$ and $d$ in order
to prevent numerical instabilities.
In particular, for NCOP, $\Delta t = 0.01$ for all values of $d$
and $N$ except when $N=10$. In this latter case,
we have taken  $\Delta t = 0.005$.
For COP, $\Delta t = 0.05$ for $d=1$ and $\Delta t = 0.01$ for $d=2$.
For the parameters of the potential $V(\vec{\phi})$ we have chosen
the values $r=-10$ and $g=1$.

After computing the structure factor $C(\vec{k},t)$
for several different times, the spectrum of scaling
exponents $\varphi(x)$ can be
obtained by using the general scaling form (1.10), which can be rewritten as
\begin{equation}
\ln C(\vec{k},t) = \varphi(x) \ln L(t) + \ln F(x)
\end{equation}
where $L(t) = t^{1 \over 2+p}$ and $x= kL(t)$.
Hence, plotting $\ln C(x/L(t),t)$ vs. $\ln L(t)$ at fixed $x$ one can measure
$\varphi(x)$ from the slope and $F(x)$ from the intercept with the vertical
axis.
However, from the numerical point of view it is more convenient to use a
slightly different procedure, because $x/L(t)$ could
turn out to be too small
or too big with respect to the available values of $k$.
For the NCOP case, $C(\vec{k},t)$ is computed not as $C(x/L(t),t)$ but as
$C(x k_2(t),t)$ where $k_2(t)$ is defined by
$C(k_2(t),t) = C(0,t)/2$.
This introduces in Eq. (3.1) an additional constant term given by the
logarithm of the proportionality factor between $L(t)$ and $k_2(t)$.
Furthermore, the error in the determination of $k_2(t)$ and $C(x k_2(t),t)$
is greatly reduced by the use of linear interpolation between the discrete
values of $k$.
In the COP case $C(\vec{k},t)$ is computed as $C(x k_m(t),t)$ where $k_m(t)$
is the peak wave vector of the structure factor. The logarithm of
$C(x k_m(t),t)$ is plotted versus $\log(L^2(t) k^{2-d}_m(t))$.
With this choice, for $N=\infty$,  the slope $\varphi(x)$
is given by $d \psi(x)$
rather than by $d \psi(x/u)$. This makes the comparison between
numerical and analytical results easier also for $N<\infty$.
Again the quality of the fit is enhanced by determining the peak wave vector
and $C(x k_m(t),t)$ via cubic and linear interpolation, respectively.

In order to check the quality of the numerical method, let us
consider the $N=\infty$ case where exact analytical results
are available. We solve for $C(\vec{k},t)$ in one and two
space dimensions with $\Delta$ ranging in the interval $(0.01, 10)$. We
discuss first the case of NCOP and then the case with COP.
The motivation for doing this computation is also to establish
clearly the behaviour of observables according to standard
scaling (NCOP) and to multiscaling (COP).

\vspace{5mm}

\noindent{\it NCOP}

\noindent In order to analyse the behaviour of $Y(t)$ in
Fig.1  $\ln Y(t)$ has been plotted versus
$\ln (\ln t)$ for $d=1$ and $d=2$. In both cases $\ln Y(t)$
displays the approach to the asymptotic constant value
$\ln d/4$ of standard scaling predicted by (2.15)
through a transient dependent
on the initial condition $\Delta$. No detectable dependence on
the initial condition is found in the behaviour of $\varphi(x)$
which in agreement with (2.6) follows the constant behavior $\varphi(x)
\equiv d$. Similarly, the numerical results for the
scaling function reproduce accurately the gaussian behavior (2.11).

\vspace{5mm}

\noindent {\it COP}

\noindent With COP the behaviour of $\ln Y(t)$ is qualitatively
different from what we had above with NCOP. In place of the
relaxation to a constant value now (Fig.2)
there is an upward increasing trend revealing multiscaling.
In the time of the computation there is still a dependence
on the initial condition $\Delta$, with a faster convergence to the
asymptotic behaviour $\sim \ln (1/2\ln (t))$ given by
(2.15) for higher values of $\Delta$. The
asymptotic behaviour has not been reached in the time
of the computation due to the much slower dynamics
of COP.

Multiscaling is most clearly illustrated by the behaviour of $\varphi(x)$.
It is interesting to see how the spectrum of exponents depends on the
time interval of observation.
In Fig.3 the evolution of $\varphi(x)$
in subsequent time intervals has been plotted for different
values of $\Delta$ and for $d=1$. Results for $d=2$ are
similar. Fig.3 demonstrates the relaxation of $\varphi(x)$
to the asymptotic behaviour given by $\varphi(x) =
d \psi(x)$. As remarked above the relaxation is faster
for higher values of $\Delta$. The late stage results are
displayed in Fig.4 both for $d=1$ and $d=2$ showing the independence
from $\Delta$ of the computed $\varphi(x)$. This suggests that, at least
in the range of $x$ considered, $\varphi(x)$ reaches the asymptotic regime
faster than $Y(t)$.

\vspace{8mm}

\setcounter{chapter}{4}
\setcounter{equation}{0}

\section*{4 - Results for finite $N$}

\vspace{5mm}

In this section we illustrate the solution of the BH equation with
finite $N$ obtained by the numerical method described in the
previous section.

\vspace{5mm}

\noindent {\it NCOP}

\noindent The standard scaling behaviour of systems with
NCOP is manifested (Fig.5)
first of all in the behaviour of $ Y(t)$ which
according to (2.15) goes to a constant asymptotic value $b(N)$
smaller then $d/4$ and decreasing monotonically with $N$.
The transient preceding the asymptotic behaviour now depends both on $\Delta$
and $N$. Asymptotic behaviour independent of $\Delta$ and $N$ instead
is manifested by $\varphi(x)$ which displays with great
accuracy the flat behaviour $\varphi(x) \equiv d$. A significant $N$
dependence shows up (Fig.6) however in the scaling function $F(x)$.
According to the analysis of section 2, a deviation from gaussian
behaviour is expected for finite $N$ in the tail of $F(x)$ due to the
second term in the right hand side of (2.5) and this deviation
clearly should be more important for small values of $N$. The plot of $\ln
F(x)$
vs. $x$ reveals the interesting feature that the tail decays exponentially
rather than following the generalised Porod's law
$\sim x^{-(d+N)}$[14]. Simulations of a system with NCOP and $N>d$ have been
performed by Toyoki[15] finding a tail which decays with a power much higher
than that of the generalised Porod's law. Our result suggests
that an exponential fit might be appropriate also in this case.

\vspace{5mm}

\noindent {\it COP}

\noindent The picture is more complex and interesting with COP.
Fig.7 and Fig.8 display respectively the behaviour of $\ln Y(t)$
for a fixed value of $N$ with varying $\Delta$ and viceversa
for a fixed value of $\Delta$ with varying
$N$. What emerges from a comparison with the analogous
data for $N=\infty$ is that for fixed $N$ (here $N=300$)
the behaviour is
of the standard scaling type
for $\Delta$ sufficiently small (e.g $\Delta=10^{-5}$) while
it is of the multiscaling type for $\Delta$ large ($\Delta=10$)
with an interpolating behaviour for intermediate values of $\Delta$.
Similarly, for $\Delta$ fixed ($\Delta =0.01$) the behaviour goes
from standard scaling for $N=300$ toward multiscaling as $N$ grows very
large. This pattern fits with the crossover picture illustrated
in the Introduction allowing for a crossover time $t^{\ast}$
which grows both with $N$ and with $\Delta$. Standard
scaling then applies when the values of $N$
and $\Delta$ are such that $t^{\ast}$ is very short. For higher
values of $N$ and $\Delta$, instead, $t^{\ast}$ can be made long
enough for the system to develop
multiscaling behaviour before
the asymptotic standard scaling behaviour is reached.
If only multiscaling behaviour is observed,
as for instance for $N=300$ and $\Delta=10$ or for $N=10^{6}$
and $\Delta=0.01$, it means that for those values of $N$ and
$\Delta$ the crossover time $t^{\ast}$ is larger than the
maximum time reached in the numerical computation. From these data
it is very difficult, though, to infer the quantitative dependence of
$t^{\ast}$
on $N$ and $\Delta$. BH have proposed[8] the analytical form
$t^{\ast} \sim (\Delta N)^{4/d}(\ln N)^{3}$,
which holds for $d>1$. However, the
predictions from this formula seem to be quite off from what
we observe. For instance for $N=300, \Delta=1$ and $d=2$ the
above formula gives $t^{\ast} \sim 10^{7}$ which is large enough to
expect an observable crossover, while for these values of the
parameters we find only standard scaling behaviour (Fig.7 and Fig.9).

As we have seen with $N=\infty$ the distinction between standard
scaling and multiscaling is most effectively manifested through
$\varphi(x)$. Thus, according to the crossover picture obtained
from $Y(t)$, it should be possible to produce
standard scaling or multiscaling
in $\varphi(x)$ by properly choosing the values of
$N$ and $\Delta$.
In Fig.9 we have analysed the
evolution of $\varphi(x)$ in time for $\Delta=0.01$ and
different values of $N$ for $d=2$ (similar results are
obtained for $d=1$). For $N$ ranging from $10^{2}$ to
$10^{4}$, $\varphi(x)$ displays standard scaling over all time intervals
implying that $t^{\ast}$ is of order one. For larger values $N=10^{5},
10^{6}$ one can definitely recognize multiscaling type of
behaviour over the initial time intervals evolving toward
standard scaling in the later time intervals. Here it is
difficult to assess the value of $t^{\ast}$, but it must be
of the order of magnitude of the time of observation. Finally,
for $N=10^{7}$ the behaviour of $\varphi(x)$ is of the
multiscaling type over all time intervals implying that $t^{\ast}$ is
larger of the time of computation.

In order to complete the analysis in Fig. 10 we have plotted the
logarithm of the scaling function $F(x)$ vs. $x$ for different values
of $N$ finding again an exponential tail like in the NCOP case.
In this case there are small secondary peaks superimposed on the tail
which scale like $L(t)$ and which become more pronounced as
$N$ grows. Exponential tails have been observed previously
in the simulation of systems with COP and without topological
defects[7].
Finally, in agreement with Rojas and Bray[12]
we find that the peak of $F(x)$ is well fitted
by the quartic exponential form appearing in the BH analytical solution.

\vspace{8mm}

\setcounter{chapter}{5}
\setcounter{equation}{0}

\section*{5 - Conclusions}

\vspace{5mm}

The main motivation for this paper was to investigate in detail the
onset of standard scaling in the BH model for phase ordering
kinetics with COP and finite $N$. We have done this by a
comparative study of the numerical solution of the model
with NCOP and with COP. In both cases eventually there is
standard scaling, but the difference is much more profound
than just the value of the growth exponent ($z=2$ for NCOP and
$z=4$ for COP) when the whole development of the dynamics is
taken into account. As probes for scaling we have used $Y(t)$ and
$\varphi(x)$. Parameters of the quench are the initial condition
$\Delta$ and the number of components $N$ of the order parameter.

The picture for NCOP is the following. Starting from a uniform
initial condition $C(\vec{k},t=0)=\Delta$, after a short
transient of duration $t_{0}$
during which information on the initial condition
is lost, the dynamics of standard scaling sets
in, with ordered regions growing like $L(t) \sim t^{1/2}$,
$\varphi(x) \equiv d$ and the scaling form (2.6) obeyed.
The only place where there remains a detectable transient
dependence on the initial condition $\Delta$ for longer times than
$t_{0}$ is in the behaviour of $Y(t)$ which displays a very slow
approach to the constant asymptotic behaviour. This means that
in the scaling ansatz for $R(t)$ there is a slow correction
with a small amplitude. This pattern of behaviour for NCOP
is the same for any value of $N$, including $N=\infty$.

By contrast, with COP the way the system eventually reaches
standard scaling is more complicated and
depends on $N$ due to the existence of the two
characteristic times $t_{0}$ and $t^{\ast}$.
Only if  $t^{\ast} \sim t_{0}$ there is no observable difference
between COP and NCOP. Instead, if $t^{\ast}$ is sufficiently
larger than $t_{0}$ the system displays multiscaling in between
$t_{0}$ and $t^{\ast}$, before the standard scaling regime
is reached. In this sense multiscaling is not only a feature of the
special case $N=\infty$, but is relevant also for systems
with finite $N$.
The existence of a connection between multiscaling and standard scaling is
expected after recognizing that these are the two asymptotic features
of a crossover process. More specifically, let us consider a
general scaling form containing both regimes
\begin{equation}
C(\vec{k},t,N)=L^{d\psi(x)} {\cal F}(x,\frac{N}{L^{a}})
\end{equation}
with $x=k/k_{m}$, $a$ an index to be determined
and ${\cal F}$ a function with the limiting behaviors
\begin{equation}
{\cal F}(x,\frac{N}{L^{a}})= \left\{ \begin{array}{ll}
                       1       & \mbox{if $N/L^{a} \gg 1$} \\
                       A(\frac{N}{L^{a}})^{\alpha(x)}   &
\mbox{if $N/L^{a} \ll 1$}
\end{array}
\right.
\end{equation}
where $A$ and $\alpha(x)$ also must be determined. The above form clearly
yields multiscaling if the limit $N \rightarrow \infty$ is taken.
If instead $N$ is kept finite and $L(t)$ gets large one has
\begin{equation}
C(\vec{k},t,N)=AL^{d\psi(x)-a\alpha(x)} N^{\alpha(x)}
\end{equation}
and imposing $d\psi(x)-a\alpha(x)=d$ one finds standard
scaling
\begin{equation}
C(\vec{k},t,N)=AL^{d}e^{-\frac{d}{a}(1-\psi(x))\ln N}=
AL^{d}e^{-\frac{d}{a}(x^{2}-1)^{2}\ln N}
\end{equation}
exactly with the BH scaling function revealing the deep connection
between multiscaling and standard scaling as the
multiscaling spectrum $\psi(x)$ dictates the form of
the scaling function in the standard scaling regime.
This multiscaling to standard scaling crossover in principle could be
observed also in systems with realistic values of $N$ by
making $t^{\ast}$ large enough exploiting the dependence of
$t^{\ast}$ on $\Delta$. In this respect it might be interesting
to check this hypothesis on the simulations of ref.s [5,6,7]
performed with values of $\Delta$ making $t^{\ast}$ sufficiently large.

Finally, the finding of exponential tails in the scaling functions
is quite interesting and CDS simulations are under way in order to
check on the existence of these tails in systems with $N>d+1$.

\vspace{2cm}

{\bf Acknowledgements}

\noindent It is a pleasure to thank Antonio Coniglio and Sharon Glotzer
for the many clarifying and quite instructive discussions on the subject of
this work. We wish also to thank the referee for suggestions which
have led to an improvement of the paper.
One of us (C.C.) wishes to thank the Polymers Division of
the National  Institute of Standards and Technology for hospitality
during completion of this work and the Boston University Center
for Computational Science for generous use of their computing
facilities.

\newpage

{}~~\\
{}~~\\
{\bf  References}
{}~~\\
\begin{enumerate}

\item For a recent review see A.J.Bray, {\it Adv. Phys.}
{\bf 43}, 357, (1994)

\item G.F.Mazenko and M.Zannetti, {\it Phys. Rev. Lett.}
{\bf 53}, 2106, (1984)

\noindent A.J.Bray, {\it J. Phys. A} {\bf 22}, L67, (1990)

\noindent J.G.Amar and F.Family, {\it Phys. Rev. A} {\bf 41},
3258, (1990)

\noindent T.J.Newman, A.J.Bray and M.A.Moore, {\it Phys.
Rev. B} {\bf 42}, 4514, (1990)

\item A.Coniglio and M.Zannetti, {\it Europhys. Lett.} {\bf 10},
575, (1989)

\item F.Liu and G.F.Mazenko, {\it Phys. Rev. B} {\bf 45}, 6989, (1992)

\item A.Coniglio, Y.Oono, A.Shinozaki and M.Zannetti,
{\it Europhys. Lett.} {\bf 18}, 59, (1992)

\item M.Mondello and N.Goldenfeld, {\it Phys. Rev. E} {\bf 47},
2384, (1993)

\noindent M.Siegert and M.Rao, {\it Phys. Rev. Lett.}
{\bf 70}, 1956, (1993)

\item M.Rao and A.Chakrabarti, {\it Phys. Rev. E} {\bf 49},
3727, (1994)

\item A.J.Bray and K.Humayun, {\it Phys. Rev. Lett.} {\bf 68},
1559, (1992)

\item Y.Oono, private communication to M.Z.

\item C.Yeung, Y.Oono and A.Shinozaki, {\it Phys. Rev. E}
{\bf 49}, 2693, (1994)

\noindent S.De Siena and M.Zannetti, {\it
Phys. Rev. E} {\bf 50}, 2621, (1994)

\item T.J.Newman and A.J.Bray, {\it J. Phys. A} {\bf 23},
4491, (1990)

\item F.Rojas and A.J.Bray, {\it Phys. Rev. E} {\bf 51},
188, (1995)

\item G.F.Mazenko, {\it Phys. Rev. Lett.} {\bf 63}, 1605, (1989);
{\it Phys. Rev. B} {\bf 42}, 4487, (1990) and
{\it ibidem} {\bf 43}, 5747, (1991)

\item A.J.Bray and S.Puri, {\it Phys. Rev. Lett.} {\bf 67},
2670, (1991)

\noindent H.Toyoki, {\it Phys. Rev. B} {\bf 45}, 1965, (1992)

\item H.Toyoki, {\it Mod. Phys. Lett. B} {\bf 7}, 397, (1993)

\end{enumerate}

\newpage

{\bf Figure Captions}

\noindent Fig.1 - $\ln Y(t)$ for NCOP with $N= \infty$ and different
values of $\Delta$ displaying the approach to $\ln(1/4)=-1.38629$
for $d=1$ and to $\ln(1/2)=-0.69314$ for $d=2$ as predicted by (2.15).

\noindent Fig.2 - $\ln Y(t)$ for COP with $N=\infty$ and different values
of $\Delta$.

\noindent Fig.3 - Time evolution of $\varphi(x)$ for COP with
$N=\infty$, $d=1$ and different values of $\Delta$.

\noindent Fig.4 - Late stage $(200000<t<500000)$ multiscaling
behaviour of $\varphi(x)$ for COP with $N=\infty$ revealing independence
from the initial condition.

\noindent Fig.5 - $\ln Y(t)$ for NCOP with $N=10,10^{3},10^{6}$.
The $\Delta$ dependent transient preceding the asymptotic
behaviour is negligible in this scale for $d=1$.

\noindent Fig.6 - Plot of the scaling function for NCOP
demonstrating exponential decay in the tails.

\noindent Fig.7 - $\ln Y(t)$ for COP with $N=300$ displaying,
in the preasymptotic regime, the switch from standard scaling to
multiscaling with increasing $\Delta$.

\noindent Fig.8 - $\ln Y(t)$ for COP with $\Delta=0.01$
displaying, in the preasymptotic regime, the switch from standard
scaling to multiscaling with increasing $N$.

\noindent Fig.9 - The evolution of $\varphi(x)$ for COP with $\Delta
=0.01, d=2$ and different values of $N$.

\noindent Fig.10 - Plot of the scaling function for COP demonstrating
exponential decay in the tails. Computations have been carried out
with $\Delta =0.01$.

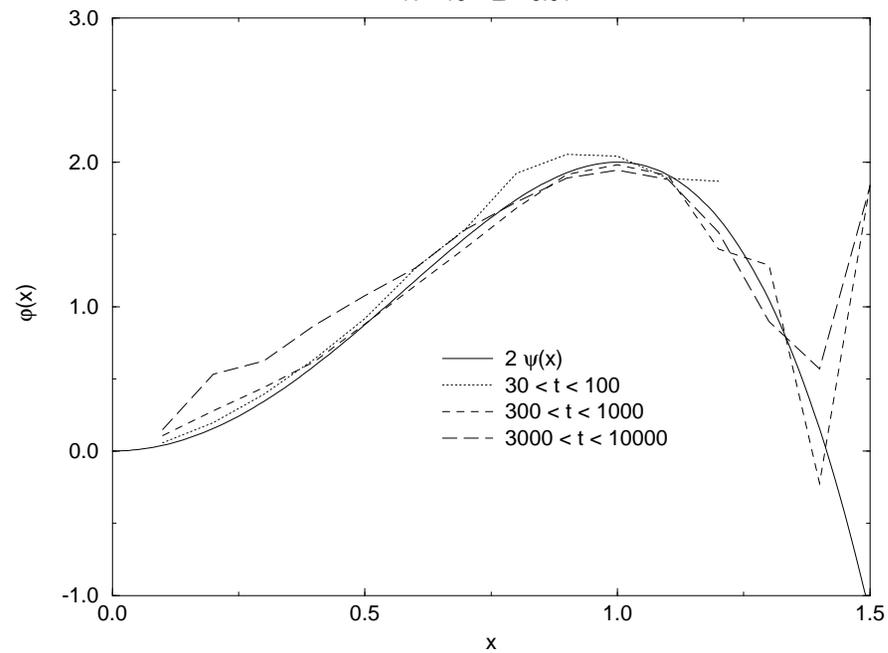
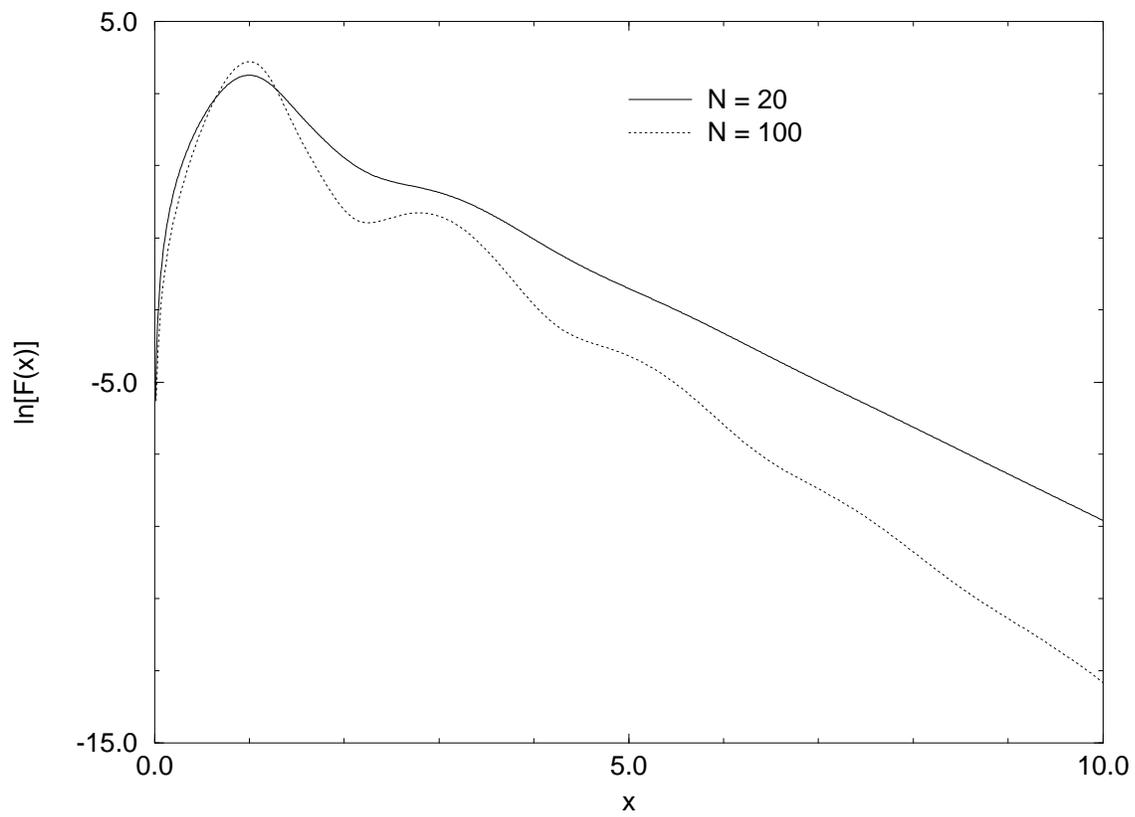
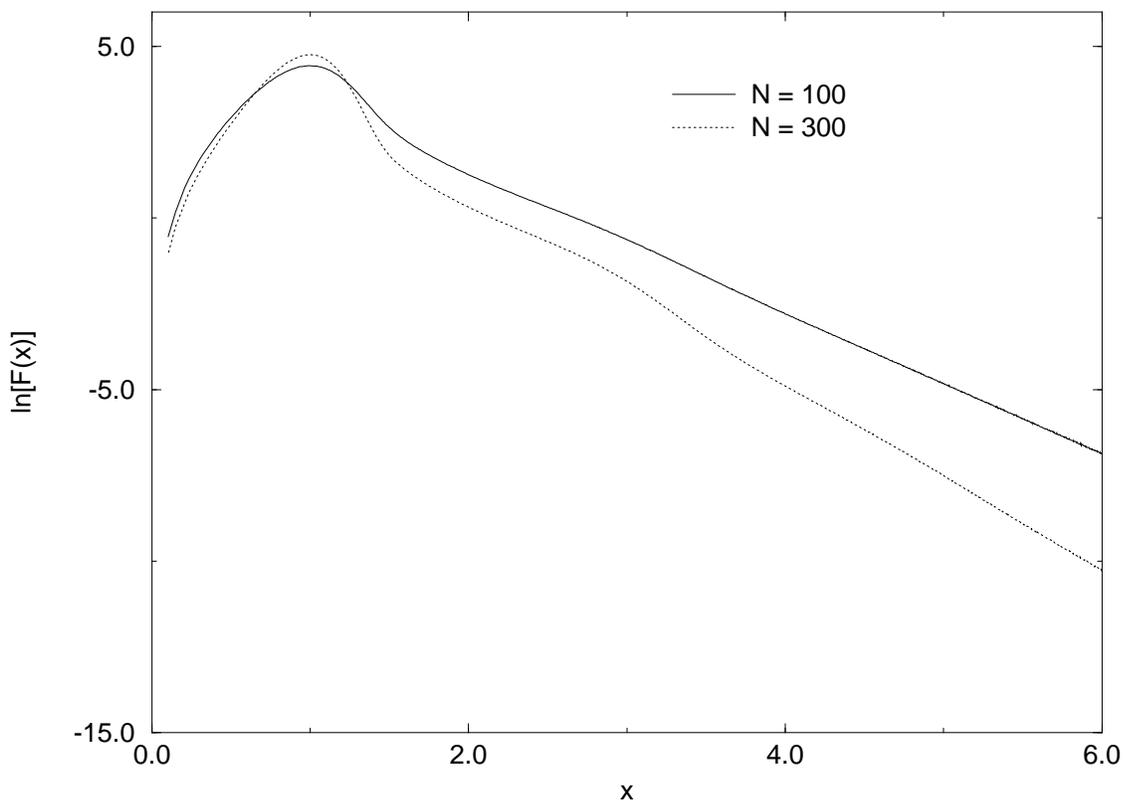
\end{document}